\documentclass[fleqn]{article}
\usepackage{graphics}

\usepackage{latexsym}

\begin{document}
\baselineskip 8truemm
\begin{titlepage}
\vspace{5mm}
\begin{center}
{\Large {\bf Boson ground state fields in electroweak theory with non-zero
charge densities} }
\vspace{6mm} \\
{\bf J. Syska}\footnote{jacek@server.phys.us.edu.pl \\
{\normalsize Special thanks to Marek Biesiada for the correction of the text.}}\\
\vspace{5mm}
{\sl Modelling Research Institute, Drzyma{\l}y 7 m.5, 40-059 Katowice, Poland}\\
{\sl and Department of Field Theory and Particle Physics, Institute of Physics,} \\
{\sl University of Silesia, Uniwersytecka 4, 40-007 Katowice, Poland}\\
\noindent

\setcounter{equation}{0}
\vspace{5 mm}
ABSTRACT
\end{center}
\vspace{2 mm}

The "non-linear" self-consistent theory of classical fields in the electroweak model
is proposed. Homogeneous boson ground state solutions in the GSW model at the presence
of a non-zero extended fermionic charge densities are reviewed and fully reinterpreted
to make the theory with non-zero charge densities \cite{JacekManka} fruitful.
Consequences of charge density fluctuations are proposed.

\vspace{20 mm}

PACS No. 03.65.Sq

\vspace{2 mm}
\vfill
\vspace{5 mm}
\end{titlepage}

\section{Introduction}

In quantum field theory matter particle is treated as a set of
strictly point-like aggregated quanta. Even in classical field
theory this creates unpleasant problems such like the infinite
self-interaction energy of a point charge, for example. In
quantum theory, these zero-point energy divergences do not
disappear; on the contrary, they are getting worse, and despite
the comparative success of renormalization theory the feeling
remains that there ought to be a more elegant way of doing
things. Yet it is not only a problem of the aesthetics. Even if in
times of Feynman and Dirac more physicists viewed renormalization
ideas as mathematically illegitimate (suspect at least)
\cite{Ruger} than they used to do today. Even if a certain theory
had a fantastic calculational results being in agreement with
today's experiments, one should remember that one is only able to
falsify physical theories\footnote{{\it One is only able to
falsify physical theories and not to prove them directly}: This is
in fact a definition of empirical science, recognized since Popper
\cite{Popper}.}. There exists also a more practical problem:
namely the particular model of quantum field theory is
satisfactory only if the renormalization sequence of
approximations converges. What we now know is merely that in few
of the theories the first few terms give good agreement with
experiment. Finally, from the theoretical point of view,
particles are not treated in the even way. What do I mean by
this? When in quantum mechanics the one-particle probabilistic
interpretation of the wave-function $\phi$ for a particle is
forced, such interpretation causes a problem for one kind of them
(e.g. scalar particles) whereas for others (for example, fermion
particles) it causes none. For the Klein-Gordon equation for
example, it resulted not only in the rejection of the
probabilistic interpretation of the wave-function (in such
circumstances this action would sound reasonably) but in the
persuasion that the "interpretation of the Klein-Gordon equation
as {\it single-particle equation}, with wave-function $\phi$,
therefore also has to be abandoned" \cite{Ryder}. Yet, when
quantum field theory enters, the same functions are
"re-interpreted as the charged density" \cite{Ryder}, and we
wonder why, in quantum mechanics, this interpretation is not
commonly allowed and instead the giving up of the Klein-Gordon
equation is forced. For fermion particles this problem does not
appear. Even in the body of quantum mechanics there are
statements which are clearly inconsistent. For example when the
scalar wave-function is complex, only probabilistic
interpretation is considered as valid but for the scalar
wave-function which is real, "$\phi$ corresponds to electrically
neutral particles, and $\rho$ and ${\bf j}$ are then the charge
and current densities, rather than the probability and probability
current densities." \cite {Ryder}.

On the other hand, non-linear field theories possess extended
solutions, referred to as solitons, which represent stable
configurations with a well-defined finite energy. Since
non-Abelian gauge theories are non-linear it is natural to
suspect that this new type of solutions may be of relevance to
particle physics. Indeed the last ten years was a period of
intensive studies of vortices, magnetic monopoles and
'instantons', which are all specific types of solitonic solutions.
If gauge theories are taken seriously, so must these solutions be
taken too. A question arises: do they give rise to a new physics
\cite{Barut} ?

Some years ago one suggested that the electromagnetic field in
the presence of external charge is unstable \cite{bib
Gerschtein}. In effect a new charged non-standard ground state
accompanied by particle-antiparticle pairs might appear. In their
paper \cite{bib Muller} M\"{u}ller, Rafelski and Greiner wrote:
"{\it The question arises, whether q.e.d. }(quantum
electrodynamics) {\it of strong fields is an ensemble of academic
problems or whether it can be subject to experimental tests. We
believe, that the basic new phenomenon of positron autoionization
can be experimentally verified in heavy ion collisions. In the
long future even $\gamma$-transitions of the superheavy
intermediate molecules - though collision broadened - may be
observed and may lead to further tests of the theory. It is
encouraging to have recent reports by Armbruster and Mokler and
by Saris et al.\footnote{ Mokler, P.H., Stein, H.J., Armbruster,
P.: Contribution to the Atlanta-Conference on Atomic
Spectroscopy, Atlanta (1972). Mokler, P.H.: GSI-Bericht 72-10.
See also: Saris, F.W., Mitchell, I.V., Sandy, D.C., Davies, J.A.,
Laubert, R.: Radiative transitions between transient molicular
orbitals in atomic collisions. Contribution to the
Atlanta-Conference on Atomic Spectroscopy, Atlanta (1972). } on
first experimental findings in this direction. Both effects, the
positron autoionization and espacially the $\gamma$-transitions
in superheavy quasimolecules can possibly - in the long - be also
observed with higher resolution by going through nuclear compound
states with life time of the order $10^{-17} - 10^{-18}$ sec. }"
Similarly in non-Abelian theories a ground state of boson fields
induced by the external (non-bosonic) charge may change the
physical system. We may expect the appearance of such ground
state boson field configurations in very dense microscopic
objects created in heavy ion collisions \cite{bib Muller}. Ground
state configurations of boson fields are also the subject of
interest in the  field of astrophysics where the presence of
superdense matter is expected in massive compact objects like for
example, neutron stars or even more exotic case\cite{bib
Rafelski}.

The aim of this paper is to examine the phenomenon of homogeneous
boson ground state solutions (boson ground fields) in the
Glashow-Salam-Weinberg (GSW) theory \cite{JacekManka}. Because
the Higgs scalar has not been found and the real symmetry breaking
mechanism has not been confirmed, I use in this paper the notion
of the Higgs scalar in the context of the weakly charged
galactical ether\footnote{special thanks to Karol Ko{\l}odziej
for this notion} in which we are immersed.

\section{The General Theory}

The mathematics of non-linear Schr\"{o}dinger and Dirac equations is quite
different from that of linear equations.
The Hilbert space formulation of quantum theory owes its origin to linearity of
the Schr{\"o}dinger equation.
Consequently in non-linear theories the Hilbert space formulation calls for
modification.
It may work approximately if the non-linear terms are small and treated as
perturbations. But it is not always the case.

The "non-linear" self-consistent classical field theory\footnote{
This interpretation is connected with the existence of the self
field. The self field is small for atomic phenomena, but it is
not always small. There is another region where the non-linear
term dominates \cite{Barut}. In perturbative QED the self field
of the electron is completely absent and it comes back in via a
separate quantized radiation field "photon by photon", whereas in
the self-consistent classical field concept the whole self field
has been put in from the beginning. }, a non-abelian case of
which is proposed below, has been previously used with great
success in the abelian case by Barut, Kraus, Van~Huele, Dowling,
Salamin, \"{U}nal \cite{bib B-K-1}, \cite{bib B-H}, \cite{bib
B-D}, \cite{spontaneous}, \cite{Unal} (see also \cite{bib J},
\cite{bib M}).

But here a serious warning has to be given. In Barut coupled equations the
wavefunction $\Psi(x)$ has not the interpretation connected with the full
charge density distribution, as in the original linear Schr\"{o}dinger or
Dirac equations, but it is connected with the fluctuations of charge density
distribution.
Hence that model has its ambition to give the results which till now were
attributed to QED, even overcrossing its applicability.

In the model the covariant differentiations, $D_{\mu}$ for the Higgs doublet
$H$, and $\nabla_{\mu}$ for a fermionic field $R$, are
\begin{equation}
\label{Pdz_5}
D_{\mu}H = \partial_{\mu}H + igW_{\mu}H + \frac{1}{2}ig'YB_{\mu}H \;\; ,
\end{equation}
\begin{equation}
\label{Pdz_7}
\nabla_{\mu}R = \partial_{\mu}R + \frac{1}{2}ig'YB_{\mu}R \; ,
\end{equation}
where
\begin{equation}
\label{Pdz_8}
W_{\mu} = W_{\mu}^{a} \frac{\sigma^{a}}{2}
\end{equation}
is the gauge field decomposition with respect to the $su(2)$ algebra
generators.

The $U_{Y}(1)$ field tensor is defined as
\begin{equation}
\label{Pdz_3}
B_{\mu \nu} = \partial_{\mu}B_{\nu} - \partial_{\nu}B_{\mu}
\end{equation}
and the $SU_{L}(2)$ Yang - Mills field tensor as\footnote{
Non-linear self-interactions of $W^{a}_{\nu}$ bosons have never
been proved either a priori (it's obvious) or in consequences
(so, what's the reason for their existence), hence I will count
them to be only the curiosity of the model.(?) In the presented
model the non-linear hypothetical ground state configurations of
the $W$ and $Z$ fields might have appeared when having being
induced by the interaction with the nonzero fermionic charge
density fluctuations.}
\begin{equation}
\label{Pdz_4}
F_{\mu \nu}^{a} = \partial_{\mu}W_{\nu}^{a} - \partial_{\nu}W_{\mu}^{a} -
g\varepsilon_{a b c }
W_{\mu}^{b}W^{c}_{\nu} \; ,
\end{equation}
where the $\varepsilon_{a b c}$ are the structure constants for $SU_{L}(2)$
($\varepsilon_{a  b   c}$ is antisymmetric under the interchange of two
neighbour indices and $\varepsilon_{1 2 3} = +1 $ ).
The coupling constant for $SU_{L}(2)$ is denoted by $g$, and by
convention the $U_{Y}(1)$ coupling is $g'/2$. The weak hypercharge operator
for the $U_{Y}(1)$ group is called $Y$.

Now, the "Higgs doublet":
\begin{equation}
\label{Pdz_11}
H = \frac{1}{\sqrt {2}}\left(\matrix{0\cr
                             \varphi \cr}\right)
\end{equation}
contains the fluctuation $\varphi$ of the Higgs field.

For the sake of simplicity and transparency I specify only the electron and
its neutrino. The contributions from other existing fermions can be treated in
a similar way.
Here we adopt the notation
\begin{equation}
\label{Pdz_12}
L = {\nu_{L} \choose e_{L}}\;\; {\rm and} \;\; R = (e_{R}) \; .
\end{equation}

In accord with the above statement, connected with the $\Psi(x)$
function, let us make the following formal replacement of the
uncharged (standard model) SM physical configuration (which is in
the neighbourhood of $\varphi = v$) with fields $W^{a}_{\mu}$,
$B_{\mu}$, $\varphi$ by the charged physical configuration (which
is in the neighbourhood of $\varphi = \delta$, see below) with
fields $W^{a}_{\mu}$, $B_{\mu}$, $\varphi$ (L{\small HS} of
Eq.(\ref{shifts})) which we decompose into (R{\small HS} of
Eq.(\ref{shifts})) the changing part $W^{a}_{\mu}$, $B_{\mu}$,
$\varphi$ and the part in the minimum of the effective potential
${\cal U}_{ef}$; ($v$ and $\delta$ are the different quantities;
$v$ is a background "field", $\delta$ is the fluctuation):
\begin{eqnarray}
\label{shifts}
\left\{ \begin{array}{lll}
W^{a}_{\mu} =: W^{a}_{\mu} + \omega^{a}_{\mu} \; , \\
B_{\mu} =: B_{\mu} + b_{\mu} \; ,  \\
\varphi =: \varphi+ \delta \; .
\end{array} \right.
\end{eqnarray}

Now, the "non-linear" Lagrangian density of electroweak model with hidden
$SU_L(2) \times U_Y(1)$ symmetry is:
\begin{equation} \label{lagrangian}
{\cal L} = - \frac{1}{4}\;F^a_{\mu \nu}F^{a \mu \nu} -
\frac{1}{4}\;B_{\mu\nu}B^{\mu\nu} + (D_{\mu}H)^{+}D^{\mu}H -
\lambda (H^{+} H - \frac{1}{2} v^{2})^{2} + {\cal L}_f \; ,
\end{equation}
where ${\cal L}_f$ is the fermionic part, $\lambda$ and $v$ are constant
parameters.

Field equations for the Yang-Mills fields (with still non-shifted field
$\varphi$) are following\footnote{
The formal shape of Eqs.(\ref{Pdz_14})-(\ref{Pdz_16}) would be also true for
the external boson fields penetrating just
discussed configuration of boson ground fields induced by matter sources.
}
$(\Box = \partial_{\nu} \partial^{\nu})$, for $B^{\mu}$
\begin{equation}
\label{Pdz_14}
- \Box B^{\mu} + \partial^{\mu}\partial_{\nu}B^{\nu} =
-\frac{1}{4}gg'\varphi^{2}W^{3 \mu}
+ \frac{1}{4}g'^{2}\varphi^{2}B^{\mu} - \frac{g'}{2} j_{Y}^{\mu} \; ,
\end{equation}
for $W^{a \mu} (a = 1,\; 2)$
\begin{eqnarray}
\label{Pdz_15}
- \Box W^{a \mu} & + & g\varepsilon_{a b c}W^{b \nu}\partial_{\nu}W^{c \mu} =
\\ \nonumber
& = & g^{2}(\frac{1}{4}\varphi^{2}W^{a \mu} - W^{b}_{\nu}W^{b \nu}W^{a \mu} +
W^{a \nu}W^{b}_{\nu}W^{b \mu}) - g j^{a \mu} \; ,
\end{eqnarray}
and for $W^{3 \mu}$
\begin{eqnarray}
\label{Pdz_16}
- \Box W^{3 \mu} & + & g\varepsilon_{3 b c}W^{b \nu}\partial_{\nu}W^{c \mu} =
\frac{1}{4}g^{2}\varphi^{2}W^{3 \mu} - \\ \nonumber
& - & \frac{1}{4}gg'\varphi^{2}B^{\mu} - g^{2}W^{b}_{\nu}W^{b \nu}W^{3 \mu} +
g^{2}W^{3 \nu}W^{b}_{\nu}W^{b \mu} - g j^{3 \mu} \; .
\end{eqnarray}
Here $j_{Y}^{\mu}$ and $j^{a \mu}$ are to be the continuous, extended in
space\footnote{
and not operators of quantum field theory with point like charges.
}
matter current density fluctuations which are given by the equations
(hence $L$ and $R$ fields are the wavefunctions which have not the
interpretation connected with the
full charges of $L$ and $R$ particles, as in the original linear Dirac
equations, but they are connected with the charge density distribution
fluctuations of $L$ and $R$ particles, exectly as in the Barut case):
\begin{equation}
\label{Pdz_17}
j_{Y}^{\mu} = \bar{L}\gamma^{\mu}YL + \bar{R}\gamma^{\mu}YR \; ,
\end{equation}
\begin{eqnarray}
\label{Pdz_18}
j^{a \mu} = \bar{L}\gamma^{\mu}\frac{\sigma^{a}}{2}L \; , \; {\rm where} \;\;
a = 1,2,3 \; .
\end{eqnarray}

Accordingly, the fluctuation $\varphi$ of the Higgs field
satisfies\footnote{
Because the fluctuation of the Higgs field is a fluctuation of an ether
hence $\Box \varphi = 0$.
}
\begin {eqnarray}
\label{Pdz_19}
- \Box \varphi & = & (-\frac{1}{4}g^{2}W^{a}_{\nu}W^{a \nu} -
\frac{1}{4}g'^{2}B_{\nu}B^{\nu} + \frac{1}{2} gg'W^{3}_{\nu}B^{\nu})\varphi -
\nonumber \\
& - & \lambda v^{2}\varphi + \lambda\varphi^{3} +
m\frac{\varphi}{v}(\bar{e}_{L}e_{R} + \;h.c.) \; .
\end{eqnarray}

\section{Boson ground state solutions}

Now we are interested in such a configuration of fields that the ground state
of boson fields\footnote{dipped in matter} (ground fields thereafter)
$W^{a}_{\mu}$ , $B_{\mu}$ and $\varphi$\footnote{
$\varphi$ is the fluctuation of the Higgs field
%, which has to (must not) be taken "independently"
%when the Higgs field is present.
}
(L{\small HS} of Eq.(\ref{shifts})) are constant and equal just to
(R{\small HS} of Eq.(\ref{shifts}))$\omega^{a}_{\mu}$ , $b_{\mu}$
and $\delta$ respectively:
\begin{eqnarray}
\label{Pdz_21}
\left\{ \begin{array}{lll}
W^{a}_{\mu} =  \omega^{a}_{\mu} \; , \\
B_{\mu} =  b_{\mu} \; ,  \\
\varphi =  \delta \; .
\end{array} \right.
\end{eqnarray}

One can parameterize the ground fields $\omega^{a}_{\mu}$ and $b_{\mu}$ in
the following homogeneous form \cite{JacekManka}\footnote{Ryszard Ma\'{n}ka
pointed to this direction.}:
\begin{eqnarray}
\label{Pdz_22}
\omega^{a}_{\mu} =
\left\{\begin{array}{lll}
\omega^{a}_{0} = \sigma  n^{a} \; , \\
\omega^{a}_{i} = \vartheta \varepsilon_{a i b }n^{b} \;\;
{\rm and} \;\; n^{a}n^{a} = 1 \; ,
\end{array}
\right.
\end{eqnarray}
\nopagebreak[4]
\begin{eqnarray}
\label{Pdz_23}
b_{\mu} =
\left\{\begin{array}{lll}
b_{0} = \beta \; ,\\
b_{i} = 0 \; .
\end{array}
\right.
\end{eqnarray}
In Eq.(\ref{Pdz_22}) $(n^{a})$ plays the role of the unit vector in the adjoint
representation of the Lie algebra $su(2)$. It chooses a direction for the
ground field.
It is easy to see that (no summation over an index "a")
\begin{equation}
\label{Pdz_24}
\omega^{a}_{\mu}\omega^{a \mu} = \sigma^{2} n^{a} n^{a} -
\vartheta^{2} \varepsilon_{a i b } \varepsilon_{a i b } n^{b} n^{b}  \;\;\;
{\rm and} \; \; \; b_{\mu}b^{\mu} = \beta^{2} \; .
\end{equation}

When we define the "electroweak magnetic field" as ${\cal
B}^{a}_{i} = 1/2 \varepsilon_{i j k } F^{a}_{j k }$ and the
"electroweak electric field" as ${\cal E}^{a}_{i} = F^{a}_{i 0 }$
then, in the homogeneous case $(\sigma = constant , \vartheta =
constant , \beta = constant , \; (n^{a}) = constant)$, we obtain
for $\vartheta \neq 0 $ the "electroweak magnetic ground field $
<{\cal B}^{a}_{i}>_{0}$"\footnote{ Here and afterwards $ < \,
>_{0} $ means the ground state expectation value (as it is
defined in wave mechanics) on the unit volume in the
3-dimensional space. This definition is not relativisticly
covariant (hence e.g. the local Lorentz invariance might not be
the fundamental property inside discussed (meta)stable
fluctuations, yet their diameter is $\approx \; 0.149 \;\; fm$
(see Section~4.1).} and the "electroweak electric ground field
$<{\cal E}^{a}_{i}>_{0}$" in the form
\begin{eqnarray}
\label{Pdz_25}
<{\cal B}^{a}_{i}>_{0} = - g \vartheta^{2}n^{i}n^{a} \;\;\; {\rm and}
\;\;\; <{\cal E}^{a}_{i}>_{0} =
g \sigma \vartheta (\delta_{a i } - n^{a} n^{i}) \; .
\end{eqnarray}

The effective potential of our\footnote{
paper \cite{JacekManka} is drasticly reinterpreted by the current one}
model is given as the ground state expectation
value of the Lagrangian density
\begin{equation}
\label{Pdz_20}
{\cal U}_{ef} = -<{\cal L}>_{0} \; ,
\end{equation}

So the mean matter current density fluctuations $J^{\mu}_{Y}$ and $J^{a \mu}$
are the ground state expectation values of $j_{Y}^{\mu} $ and $j^{a \mu}$
respectively (see Eqs.(\ref{Pdz_17}-\ref{Pdz_19})):
\begin{eqnarray}
\label{Pdz_27}
J^{\mu}_{Y} = (<\bar{L}\gamma^{\mu} Y L>_{0} +
<\bar{R}\gamma^{\mu} Y R>_{0}) \;\; {\rm and} \;\;  \nonumber \\*
J^{a \mu} = <\bar{L}\gamma^{\mu}\frac{\sigma^{a}}{2} L>_{0} \; .
\end{eqnarray}
$J^{\mu}_{Y}$ and $J^{a \mu}$ are the extended in space quantities.

%The field equation for the shifted Higgs field is as follows (for other fields
%the formal shape of field equations for shifted fields is the same as given
%by Eqs.(\ref{Pdz_14})-(\ref{Pdz_16})):
%\begin {eqnarray}
%\label{H-shift-field}
%\Box \varphi & = & (\frac{1}{4}g^{2}W^{a}_{\nu}W^{a \nu} +
%\frac{1}{4}g'^{2}B_{\nu}B^{\nu} - \frac{1}{2} gg'W^{3}_{\nu}B^{\nu})
%(\varphi + v) - \nonumber \\
%& - & 2 \lambda v^{2}\varphi - 3 \lambda v \varphi^{2} - \lambda \varphi^{3} -
%m\frac{\varphi + v}{v}(\bar{e}_{L}e_{R} + \;h.c.) \; ,
%\end{eqnarray}
%where the boson fields $W^{a}_{\mu}$, $B_{\mu}$ and the fluctuation $\varphi$
%lie in the neighbourhood of the ground state of the system.

We now assume that we are in the local rest coordinate system in which
\begin{equation}
\label{Pdz_29}
J^{0}_{Y} = \varrho_{Y}\;\; , \;\; J^{i}_{Y} = 0\;\; , \;\;J^{a 0} =
\varrho^{a}\;\; {\rm and} \;\;J^{a i} = 0 \; ,
\end{equation}
where $\varrho_{Y}$ and $\varrho^{a}$ are the matter charge density
fluctuations related to $U_{Y}(1)$ and $SU_{L}(2)$, respectively.
Using Eq.(\ref{Pdz_20}) with Eqs.(\ref{Pdz_21}) -- (\ref{Pdz_24})
we obtain the ground state part of the effective potential \cite{JacekManka}
for the "boson ground fields induced by matter sources" configuration
(hereafter, I will call it bgfms configuration):
\begin{eqnarray}
\label{Pdz_30}
{\cal U}_{ef}(\vartheta,\sigma,\beta,\delta) & = &
-g^{2}\sigma^{2}\vartheta^{2} +
\frac{1}{2}g^{2}\vartheta^{4} - \frac{1}{8}g^{2}\delta^{2}(\sigma^{2} -
2 \vartheta^{2}) + \\ \nonumber
& + & \frac{1}{4}gg'\delta^{2}\beta\sigma n^{3} -
\frac{1}{8}g'^{2}\delta^{2}\beta^{2} + g \varrho^{a}n^{a}\sigma +
\frac{g'}{2}\varrho_{Y}\beta + \\ \nonumber
& + & \frac{1}{4}\lambda(\delta^{2} - v^{2})^{2} \; .
\end{eqnarray}

Now from the field equations, Eqs.(\ref{Pdz_14}) -- (\ref{Pdz_18}),
we can obtain four equations:
\begin{equation}
\label{Pdz_31}
\partial_{\vartheta}{\cal U}_{ef} = \partial_{\sigma}{\cal U}_{ef} =
\partial_{\beta}{\cal U}_{ef} = \partial_{\delta}{\cal U}_{ef} = 0 \; .
\end{equation}
These equations translate into four algebraic equations for the ground fields
$\vartheta$, $\sigma$, $\beta$ and $\delta$:
\begin{equation}
\label{Pdz_32}
\left[ \;\; \frac{1}{2}\delta^{2} - 2 \sigma^{2} + 2 \vartheta^{2} \;\;
\right] \vartheta = 0 \; ,
\end{equation}
\begin{equation}
\label{Pdz_33}
- g (2 \vartheta^{2} + \frac{1}{4} \delta^2) \sigma +
\frac{1}{4} g'\delta^{2} \beta n^{3} + \varrho^{a} n^{a} = 0 \; ,
\end{equation}
\begin{equation}
\label{Pdz_34}
\frac{1}{2} (g\sigma n^{3} - g'\beta)\delta^{2} + \varrho_{Y} = 0 \; ,
\end{equation}
\begin{equation}
\label{Pdz_35}
\left[ \;\; - \frac{1}{4}g^{2}(\sigma^{2} - 2 \vartheta^{2}) +
\frac{1}{2}gg'\sigma \beta n^{3} - \frac{1}{4} g'^{2} \beta^{2} +
\lambda(\delta^{2} - v^{2}) \;\; \right] \delta = 0 \; .
\end{equation}
These equations are the sreening charge analog of the screening current
condition in electromagnetism \cite{Aitchison-Hey-bis}.

Now let us choose
\begin{eqnarray}
\label{Pdz_36}
(n^{a}) = (0,0,1) \; .
\end{eqnarray}
In this case we  have an "electroweak magnetic ground field" different
from zero $<{\cal B}^{3}_{3}>_{0} = - g \vartheta^{2}$ pointed in
the $x^{3}$ spatial direction and "electroweak electric
ground fields" different from zero $<{\cal E}^{1}_{1}>_{0} =
<{\cal E}^{2}_{2}>_{0} = g \sigma \vartheta$ pointed in
the $x^{1}$ and $x^{2}$ spatial direction respectively.

Using Eqs.(\ref{Pdz_14}) -- (\ref{Pdz_19})
%and Eq.(\ref{H-shift-field})
together with Eqs.(\ref{Pdz_21})-(\ref{Pdz_24}) and
Eq.(\ref{Pdz_36}) we obtain in the ground state of the bgfms
configuration the square masses of the boson fields as
follows\footnote{ Eqs.(\ref{Pdz_37})-(\ref{Pdz_40}) together with
Eqs.(\ref{Pdz_32})-(\ref{Pdz_35}) is the ground state screening
current condition in the electroweak analog of the
electromagnetic case. }:
\begin{equation}
\label{Pdz_37}
m_{W^{1,2}}^{2} = g^{2} (\frac{1}{4}\delta^{2} - \sigma^{2} +
\vartheta^{2}) \; ,
\end{equation}
\begin{equation}
\label{Pdz_38}
m_{W^{3}}^{2} =  g^{2} (\frac{1}{4} \delta^{2} + 2 \vartheta^{2}) \; ,
\end{equation}
\begin{equation}
\label{Pdz_39}
m_{B}^{2} = \frac{1}{4} g'^{2} \delta^{2} \; ,
\end{equation}
\begin{equation}
\label{Pdz_40}
\Delta_{\varphi}^2 = {\cal N} \times f(\sigma, \vartheta, \beta, \delta) \; ,
%\left[ 2 \lambda (v^{2} + 3 \delta^{2})
%- \frac{1}{4}g^{2}(\sigma^{2} - 2 \vartheta^{2}) +
%\frac{1}{2} gg' \sigma \beta n^{3} - \frac{1}{4} g'^{2} \beta^{2} \right] \; ,
\end{equation}
where\footnote{
When $\varphi \equiv constant$, so $\Box \varphi = 0$, as it is in the case of
the ether, then the mass of the Higgs field fluctuation is a free parameter,
$\Delta_{\varphi}$, of the model. Hence $\Delta_{\varphi}$ could not be
established in this model; therefore the Higgs field fluctuation is not an
elementary particle (?) in this model.
}
$\Delta_{\varphi}$ (or rather ${\cal N}$) is a free parameter of the model.
Eqs.(\ref{Pdz_37})-(\ref{Pdz_40}) together with
Eqs.(\ref{Pdz_32})-(\ref{Pdz_35}) is the ground state screening
current condition in the electroweak analog of the electromagnetic case
\cite{Aitchison-Hey-bis}.

Let us perform (for $\delta \neq 0$) a "rotation"\footnote{
The standard model relations between the Weinberg angle
$\Theta_{W}$, $g$ and $g'$ are following:
\begin{equation}
\label{Pdz_13}
cos\Theta_{W} = \frac{g}{\sqrt{g^{2} + g'^{2}}} \;\; {\rm and} \;\; \;
sin\Theta_{W} = \frac{g'}{\sqrt{g^2 + g'^{2}}} \; .
\end{equation}
Some quantum numbers of the electroweak $SU_{L}(2) \times U_{Y}(1)$ model are
given in Table.
}
of  $W^{3}_{\mu}$ and
$ B_{\mu} $ fields to the physical fields $ Z_{\mu} $ and $ A_{\mu} $
\begin{equation}
\label{Pdz_41}
{Z_{\mu} \choose A_{\mu}} = {{cos\Theta\;\; {-sin\Theta}}
\choose{sin\Theta\;\;\;\;\; cos\Theta}}{W^{3}_{\mu}\choose B_{\mu}}
\end{equation}
and a "rotation" of  $ \sigma $ and $\beta$
ground fields to their counterparts $\zeta$ and $\alpha$ as well as a
"rotation" of the charge density fluctuations $\varrho^{3}$ and $\varrho_{Y}$
to the corresponding physical quantities $\varrho_{Z}$ and $\varrho_{Q}$
\begin{equation}
\label{Pdz_42}
{\zeta \choose \alpha} = {{cos\Theta\;\; {-sin\Theta}}\choose{sin\Theta\;
\;\;\;\; cos\Theta}}{\sigma \choose \beta} \; ,
\end{equation}
\begin{equation}
\label{Pdz_43}
{(g/cos\Theta) \varrho_{Z} \choose (g sin\Theta) \varrho_{Q}} =
{{cos\Theta\;\;{-sin\Theta}}\choose{sin\Theta\;\;\;\;\; cos\Theta}}{(g)
\varrho^{a} n^{a} \choose (g'/2) \varrho_{Y}} \; .
\end{equation}

Now using Eqs.(\ref{Pdz_37})-(\ref{Pdz_40}) and defining the $W^{\pm}$
fields as $W^{\pm} = (W^{1} \mp i W^{2})/ \sqrt{2} $
we can rewrite the square masses of the boson fields\footnote{
Because masses (and structures) of bosons change we should denote this fact
somehow or other. Hence we reserve the notation $Z^{0}$ (with the upper index
$0$) for the boson in the Standard Model case whereas $Z$ for this boson on
the ground state of the system with non-zero matter charge density
fluctuations.
} as follows:
\begin{equation}
\label{Pdz_44}
m_{W^{\pm}}^{2} = g^{2} \left[ \;\; \frac{1}{4} \delta^{2} - (\zeta
cos\Theta + \alpha sin\Theta)^{2} + \vartheta^{2} \;\; \right] \; ,
\end{equation}
\begin{equation}
\label{Pdz_45}
m_{Z}^{2} = \frac{1}{2} \left[ \;\; m_{Z^{0} \; SM}^{2} + 2 g^{2}
\vartheta^{2} + \sqrt{(m_{Z^{0} \; SM}^{2} + 2 g^{2} \vartheta^{2})^{2} -
2(gg'\delta \vartheta)^{2}} \;\; \right] \; ,
\end{equation}
\begin{equation}
\label{Pdz_46}
m_{A}^{2} = \frac{1}{2} \left[ \;\; m_{Z^{0} \; SM}^{2} + 2 g^{2}
\vartheta^{2} - \sqrt{(m_{Z^{0} \; SM}^{2} + 2 g^{2} \vartheta^{2})^{2} -
2(gg'\delta \vartheta)^{2}}  \;\; \right] \; ,
\end{equation}
\begin{eqnarray}
\label{Pdz_47}
\Delta_{\varphi}^2 = {\cal N} \times f(\zeta, \vartheta, \alpha, \delta) \; ,
%\left[ 2 \lambda (v^{2} + 3 \delta^{2}) - \frac{1}{\delta^{2}}(m_{Z}^{2}
%\zeta^{2} - m_{A}^{2} \alpha^{2}) +  \right. \\ \nonumber
%& + & \left. 2 g^{2}(\frac{1}{\delta^{2}}(\zeta cos\Theta +
%\alpha sin\Theta)^{2} + \frac{1}{4}) \vartheta^{2} \right] \; ,
\end{eqnarray}
where $ m_{Z^{0} \; SM}^{2}$ is the standard counterpart for the boson
$Z^{\mu}$ square mass which is equal to
\begin{equation}
\label{Pdz_48}
m_{Z^{0} \; SM}^{2} = \frac{1}{4}(g^{2} + g'^{2}) \delta^{2} \; .
\end{equation}

It is illustrative to write the relations between weak isotopic charge density
fluctuation $\varrho^{3}$ (see Eq.(\ref{Pdz_29}) and
Eq.(\ref{Pdz_36})), weak hypercharge density fluctuation $\varrho_{Y},$
(below defined, Eq.(\ref{Pdz_51})) standard\footnote{
unscreened
}
electric charge density fluctuation $\varrho_{Q \;SM}$,
(below defined, Eq.(\ref{Pdz_51})) standard\footnote{
unscreened
} weak charge density fluctuation $\varrho_{Z^{0} \; SM}$ and their
generalizations in our model i.e. the electric charge density fluctuation
$\varrho_{Q}$ and weak charge density fluctuation $\varrho_{Z}$:
\begin{equation}
\label{Pdz_49}
\varrho_{Q} = \varrho_{Q \; SM } + \frac{1}{2} (\frac{g'}{g} ctg\Theta - 1)
\varrho_{Y} \; ,
\end{equation}
\begin{equation}
\label{Pdz_50}
\varrho_{Z} = \varrho^{3} - \varrho_{Q} \; sin^{2}\Theta \; ,
\end{equation}
\begin{equation}
\label{Pdz_51}
\varrho_{Q \; SM } = \varrho^{3} + \frac{1}{2} \varrho_{Y} \;\; and \;\;
\varrho_{Z^{0} \; SM } = \varrho^{3} - \varrho_{Q \; SM } \;
sin^{2}\Theta_{W} \; .
\end{equation}
Here the $\Theta$ angle is the modified mixing angle given by the
formula
\begin{equation}
\label{Pdz_52}
\!\!\!\!\!\!\!\!\!\!tg\Theta = \left[ \;\; \frac{-
(1 + 8 (\vartheta/\delta)^{2})g^{2} + g'^{2}}{2 g g'} + \sqrt{(
\frac{(1 + 8 (\vartheta/\delta)^{2})g^{2} - g'^{2}}{2 g g'})^{2}
+ 1 } \;\; \right] \; .
\end{equation}
It is not difficult to see that electroweak assumptions are formally recovered
in the limit $\vartheta \longrightarrow 0$ (and extended $Q = 0$).
It is evident from Eq.(\ref{Pdz_52}) that transition from zero charge density
fluctuations to $\varrho^{3} \neq 0$, $\varrho_{Y} \neq 0$
is associated with a non-linear response of the system.

\section{Results}
\label{discus}

The calculations below are done for the boson fields in extrema of the
effective potential ${\cal U}_{ef}$.
It is not difficult to see that the solutions of
Eqs.(\ref{Pdz_32})-(\ref{Pdz_35}) for boson ground fields in the extrema of
the effective potential ${\cal U}_{ef}$ split into cases discussed just below.

\subsection{Ground fields $\vartheta \neq 0 \; $ and $ \; \delta \neq 0$ }
\label{electric}

Eqs.(\ref{Pdz_32})-(\ref{Pdz_35}) can be now rewritten as follows:
\begin{equation}
\label{Pdz_53}
\sigma = \frac{1}{2 g \vartheta^{2} } \varrho_{Q \; SM } \; ,
\end{equation}
\begin{equation}
\label{Pdz_54}
\beta = \frac{1}{g'}(g \sigma n^{3} + 2 \frac{\varrho_{Y}}{\delta^{2}}) \; ,
\end{equation}
\begin{equation}
\label{Pdz_55}
\vartheta^{6} + \frac{1}{4} \delta^{2} \vartheta^{4} - \frac{1}{4 g^{2}}
\varrho_{Q \; SM }^{2} = 0 \; ,
\end{equation}
\begin{equation}
\label{Pdz_56}
\delta^{6} + (\frac{g^{2}}{2 \lambda} \vartheta^{2} - v^{2}) \delta^{4} -
\frac{1}{\lambda} \varrho_{Y}^{2} = 0 \; .
\end{equation}
From Eq.(\ref{Pdz_55}) we see that the ground field $\vartheta$ is non-zero
only if $\varrho_{Q \; SM} \neq 0$ .

Let us notice that the relation between the weak hypercharge quantum number
$Y$ and the electromagnetic charge quantum number $Q$ can be written
in the form $Q = p Y/2$ (for matter fields), where the corresponding values
of $p\,(p\neq~0)$ are given in Table.
Then the relation between the weak hypercharge density fluctuation
$\varrho_{Y}$ and the standard electromagnetic charge density fluctuation
$\varrho_{Q \; SM }$ can be written in the similar form
\begin{equation}
\label{Pdz_57}
\varrho_{Q \; SM } = p \frac{\varrho_{Y}}{2} \; .
\end{equation}

Using Eq.(\ref{Pdz_57}) we solved numerically \cite{JacekManka}
Eqs.(\ref{Pdz_53})-(\ref{Pdz_56}) and we obtained the ground fields
squared $\vartheta^{2}$ and $\delta^{2}$ as the functions of
$\varrho_{Q \; SM}$ with $ p $ as a parameter.
Different values of $p$ (see Table) represent different matter fields which
could be the sources of charge densities.
The results of solving Eqs.(\ref{Pdz_53})-(\ref{Pdz_56}) for the $\alpha$
and $\zeta$ ground fields (see Eq.(\ref{Pdz_42})) and the $\vartheta$ and
$\delta$ ground fields are shown in Figure 1-Figure 4.

Now Eq.(\ref{Pdz_21}) has the form:
\begin{eqnarray}
\label{Pdz_58}
\left\{ \begin{array}{lll}
W^{\pm}_{0,3} = 0  \;\; , \;\;
& W^{\pm}_{1} =  \pm i \vartheta/ \sqrt{2} \; ,
& W^{\pm}_{2} =  \vartheta / \sqrt{2} \; , \\
Z_{i} = 0 \;\; , \;\;
& Z_{0} =  \zeta \;\;\; {\rm where}& (\zeta = \sigma cos\Theta -
\beta sin\Theta) \; , \\
A_{i} = 0 \;\; , \;\;
&A_{0} = \alpha \;\;\; {\rm where}& (\alpha = \sigma sin\Theta +
\beta cos\Theta) \; , \\
\varphi =  \delta \; . &  &
\end{array} \right.
\end{eqnarray}

The masses of the $Z^{0}$ and $A$ were calculated according to
Eqs.(\ref{Pdz_45})~and~ (\ref{Pdz_46}) and the appropriate
results are shown in Figs.5 - 6. The masses of the $W^{\pm}$
fields are, according to Eq.(\ref{Pdz_32}) and Eq.(\ref{Pdz_37})
(for the $\vartheta \neq 0$ configuration of fields), equal to
zero.

The results for the ratio $sin\Theta / sin\Theta_{W}$ (see
Eq.(\ref{Pdz_52})) and the physical charge density fluctuation
$\varrho_{Q}$ (see Eq.(\ref{Pdz_49})) for boson ground fields
given by Eqs.(\ref{Pdz_53})-(\ref{Pdz_56}) as functions of
$\varrho_{Q \; SM}$ are presented in Figure 7 and Figure 8,
respectively.

In all the figures the curves for different values of $p$
converge for relatively small values of $\varrho_{Q \; SM}$
(i.e., for values of $\varrho_{Q \; SM}$ in the range up to
values approximately $10^{3}$ times bigger than these which
correspond to matter densities in nucleon matter). In that range
of values for $\varrho_{Q \; SM}$ we have also that $\varrho_{Q}
\approx \varrho_{Q \; SM}$ (see Figure 8). For these reasons the
following calculations in that regime were done for $p = 1$ (for
other $p \neq 0$ in Table it would be the same).

The minimal energy density of the bgfms configuration ${\cal
E}_{min} (\varrho_{Q \; SM}) = {\cal U}_{ef}(\vartheta \neq 0,
\delta \neq 0)$ (see Eq.(\ref{Pdz_30})) for boson ground fields
given by Eqs.(\ref{Pdz_53})-(\ref{Pdz_56}) as functions of
$\varrho_{Q \; SM}$ is presented in Figure 9. For big value of
charge density fluctuation $\varrho_{Q \; SM}$ (i.e., for
$\varrho_{Q \; SM}$ which corresponds to matter densities
approximately $10^{3}$ times bigger than characteristic for
static nucleon matter) the minimal energy density ${\cal E}_{min}
(\varrho_{Q \; SM})$ is extremely big and is increasing very
rapidly with $\varrho_{Q \; SM}$. For example, ${\cal E}_{min}
\approx 2.9 \; 10^{178} \; GeV^{4}$ for $ \varrho_{Q \; SM}
\approx 1.3 \; 10^{7} \; GeV^{3}$.

For this reason a charged star could effectively resist against
the gravitational collapse. Even any global (in a star) electric
charge density distribution fluctuations in a collapsing
uncharged star\footnote{ Few years ago, Marek Biesiada claimed
that "irresistible" gravitational collapse should not be realized.
} could resist the gravitational force. It would be also true for
an electrically charged elementary particle if it has electroweak
structure; hence, it could be the reason that the wavefunction
(defined by the charge density distribution) does not collapse.

It is very interesting that when we investigate the function
${\cal E}_{min}(\varrho_{Q \; SM})$ more carefully then a subtle
structure emerges. It appears a "stable" (bgfms) configuration of
charge density fluctuation with $\varrho_{Q \; SM} \neq 0$ (see
Figure 9) different from that for the standard "linear" model (with
charge density fluctuation $\varrho_{Q \; SM} = 0$ and ${\cal
E}_{min}(0) = 0 \, $). The numerical calculations for the value of
the local minimum of the function ${\cal E}_{min} (\varrho_{Q \;
SM})$ reveal little dependence on the $\lambda$ parameter of the
Higgs potential (see Figure 9) and the results are following:
\begin{equation}
\label{Pdz_59}
{\cal E}_{min}(\varrho_{Q \; SM}) \approx (2.5811 \; GeV)^{4} \;\; {\rm for}
\;\; \varrho_{Q \; SM} \approx 0.5539 \; GeV^{3} \; .
\end{equation}

This bgfms configuration is separated from the static standard
model configuration (ground state of the standard model) by a
high barrier $\Delta{\cal E}_{min}$ which depends on the
$\lambda$ parameter (see Figure 9). For example when $\lambda = 1$
then $\Delta{\cal E}_{min} \approx (180 \; GeV)^{4}$. It is not
difficult to see that ${\cal E}_{min} \longrightarrow 0  $ as $
\varrho_{Q \; SM} \longrightarrow~0$ for all considered values of
$\lambda > 0$ and $p \neq 0$ (see Table).

When we notice that the mass of an electrically charged bgfms
configuration with the "radius of the charge fluctuation" $r_{q}$
is equal to $M_{q} = 4/3 \; \pi r_{q}^{3}$ {\small $\times$}
${\cal E}_{min}(\varrho_{Q\,SM})$, and that the matter global
electric charge fluctuation $q = 4/3 \;\; \pi \;r_{q}^{3} \;
\varrho_{Q \;SM}$ then from Eq.(\ref{Pdz_30}) and
Eqs.(\ref{Pdz_53})-(\ref{Pdz_56}) we obtain $M_{q}
\longrightarrow \pm q g v/2 = \pm q \times 80.13 \; GeV$ (sign
"+" for $Q > 0$, sign "-" for $Q < 0$) as $ \varrho_{Q \; SM}
\longrightarrow 0$ for all considered values of $\lambda >0$ and
$p \neq 0$ (see Table). The function $M_{q=1}(r_{q})$ is
presented in Figure 10. These configurations of fields lie only
on the $M_{q}-r_{q}$ curve ($M_{q} = \pm q M_{q=1}$). For
example, a droplet of the new bgfms configuration of fields with
charge fluctuation $q$ and described by Eq.(\ref{Pdz_59}) will
have the "radius of the charge fluctuation" $r_{q} = q^{1/3}
\times 0.149 \; fm$ (in comparison, for proton with full electric
charge $Q = 1$ its global electric charge radius $r_{Q} \approx
0.805 \; fm$) and the mass $M_{q} \approx \pm q \times 80.13 \;
GeV$. If one takes into account the mass of a fermion (fermions)
playing the role of matter source inducing boson ground fields,
then the value of the mass $M_{q}$ will change of about the order
of the mass of this part of a fermion (fermions) which is
contained in the region of the fluctuation.

In Figs.3-7 and Figs.9-10 the curves corresponding to $p$ and $
-p $ are the same.

\label{self-bgfms}
Now a few comments are in order.
From the Eqs.(\ref{Pdz_44})-(\ref{Pdz_47}), Eqs.(\ref{Pdz_53})-(\ref{Pdz_56})
and Eq.(\ref{Pdz_58}) we can notice that fields $W^{+}$ and $W^{-}$ taken
together as a pair of massive fields become (in this bgfms configuration of
fields) a kind of massless self field and a ground field which is coupled
to charged fields with charge density fluctuations $\varrho_{Q \; SM} \neq 0$
and $\varrho_{Y} \neq 0$. When these charge density fluctuations go to zero
then the $W^{+}-W^{-}$ ground field also goes to zero
$(\vartheta \rightarrow 0).$
The ground fields of $Z^{\mu}$ and $A^{\mu}$ given by $\zeta$ and
$\alpha$ (see Eq.(\ref{Pdz_58})) are nonzero even for $\varrho_{Q \; SM}
\rightarrow 0$ and $\varrho_{Y} \rightarrow 0$.

Because of the nonlinear terms in the field equations there appeared
the screening charge problem \cite{Aitchison-Hey-bis} which is very essential
in this paper.
Now, the role of the $|\psi|$ part of the wave function $\psi$ is played not
only by fluctuations of matter (fermion) fields but by fluctuations of weakly
charged scalar $\varphi$ (and global gauge ground fields $\zeta$ and $\alpha$).
% and the estimation of matter current densities will
% be taken in extrema of an effective potential of the model.
These fluctuation fields together with the global gauge ground fields form a
system characterized by wave functions $\varphi$, $Z^{\mu}$ and $A^{\mu}$
which are "macroscopic" in its spatial extension.
When, as in this case, we have the Higgs fluctuation field $\varphi$,
$Z_{0}$-ground field and $A_{0}$-ground field in the ground state of the
system with their ground state expectation values $< \varphi >_{0} =
\delta \neq 0$, $< Z_{0} >_{0} = \zeta \neq 0$ and
$< A_{0} >_{0} = \alpha \neq 0$ respectively, then in the
presence of the $W^{\mu}$ fields the electroweak force field generates
"electroweak screening charges" connected with the fact that both the basic
fermion field and Higgs field carry the nonzero charge densities.

\subsection{Ground fields $\vartheta  = 0 \; $ and $ \; \delta \neq 0$ }
\label{neutron}

Using Eqs.(\ref{Pdz_42}) -- (\ref{Pdz_43}) we can rewrite the effective
potential ${\cal U}_{ef}$ (Eq.(\ref{Pdz_30})) in a very simple form:
\begin{eqnarray}
\label{Pdz_67}
{\cal U}_{ef}(\zeta,\alpha,\delta) & = & - \frac{1}{8}(g^{2}
+ g'^{2})\delta^{2}\zeta^{2} + \varrho_{Z^{0} \; SM}\zeta + \varrho_{Q \; SM}
\alpha + \nonumber \\
& + & \frac{1}{4}\lambda(\delta^{2} - v^{2})^{2} \; .
\end{eqnarray}
Now, according to Eqs.(\ref{Pdz_31})-(\ref{Pdz_35}) we have
$\partial_{\alpha}{\cal U}_{ef} = \partial_{\zeta}{\cal U}_{ef} =
\partial_{\delta}{\cal U}_{ef} = 0 $
which yields
\begin{equation}
\label{Pdz_68}
\varrho_{Q \; SM} = 0 \; ,
\end{equation}
\begin{equation}
\label{Pdz_69}
\frac{1}{4} \sqrt{g^{2} + g'^{2}}\delta^{2}\zeta = \varrho_{Z^{0} \; SM}
\end{equation}
and
\begin{equation}
\label{Pdz_70}
\lambda(\delta^{2} - v^{2}) - \frac{1}{4}(g^{2} + g'^{2})\zeta^{2} = 0 \; .
\end{equation}
Non-zero weak charge density fluctuation $\varrho_{Z^{0} \; SM}$ leads
inevitably to a non-zero $\zeta$ ground field which implies $\delta \neq 0$.
From Eq.(\ref{Pdz_68}) and Eqs.(\ref{Pdz_49})-(\ref{Pdz_52}) we can also
notice that $\varrho_{Z} = \varrho_{Z^{0} \; SM}$ (for $\vartheta  = 0 \; $).

Now, combining Eqs.(\ref{Pdz_68}) and (\ref{Pdz_69}) with Eq.(\ref{Pdz_67}) we
obtain the ground state counterpart of the electroweak effective potential for
$\vartheta = 0$ (see Figure 11)
\begin{eqnarray}
\label{Pdz_71}
{\cal U}_{ef}(\delta, \; \varrho_{Z^{0} SM} ; \; \vartheta = 0,\;
\varrho_{Q \; SM} = 0 ) = \frac{2 \varrho_{Z^{0} SM}^{2}}{\delta^{2}}
+ \frac{1}{4}\lambda(\delta^{2} - v^{2})^{2} \; .
\end{eqnarray}

The solution of Eqs.(\ref{Pdz_69}) -- (\ref{Pdz_70}) leads to
\begin{eqnarray}
\label{Pdz_72} & \zeta(\varrho_{Z^{0} \; SM}) = \frac{2
\sqrt[3]{\lambda}}{\sqrt{g^{2} +  g'^{2}}} \times &  \\
\times &\left[ \sqrt[3]{\varrho_{Z^{0} \; SM} +
\sqrt{\varrho_{Z^{0} \; SM}^{2} + \frac{\lambda v^{6}}{27}}}  +
\sqrt[3]{\varrho_{Z^{0} \; SM} - \sqrt{\varrho_{Z^{0} \; SM}^{2}
+ \frac{\lambda v^{6}}{27}}} \; \right]& \geq 0  \nonumber
\end{eqnarray}
and
\begin{equation}
\label{Pdz_73}
\delta^{2}(\varrho_{Z^{0} \; SM}) =
\frac{4 \varrho_{Z^{0} \; SM}}{\sqrt{g^{2} + g'^{2}} \zeta} \; ,
\end{equation}
where $\zeta$ and $\delta^{2}$ are the functions of $\varrho_{Z^{0} \; SM}$
only.
It is not difficult to see that in the limit
$\varrho_{Z^{0} \; SM} \longrightarrow 0$ (implying
$\zeta \longrightarrow 0 $ and $\delta \longrightarrow v$)
the well-known electroweak configuration ($\delta = v$) with $U_{Q}(1)$
symmetry emerges.

Using Eqs.(\ref{Pdz_22})-(\ref{Pdz_23}) and Eqs.(\ref{Pdz_41})-(\ref{Pdz_42})
we can rewrite Eq.(\ref{Pdz_21}) for the physical field $A_{\mu}$ in the form
\begin{equation}
\label{Pdz_74}
A_{\mu} = (\alpha,0,0,0) \; .
\end{equation}
Let us notice from Eqs.(\ref{Pdz_68})-(\ref{Pdz_71}) that $ \alpha$
is not a dynamical parameter so a transformation $0 \rightarrow \alpha$
acquires the interpretation of a gauge transformation.
Here the $\alpha$ ground field corresponds to a nonphysical degree of freedom
(this is connected with the fact that $\varrho_{Q SM} = 0$) and it can be
removed by an appropriate gauge transformation. So the $U_{Q}(1)$ group
remains untouched and it gives us
\begin{equation}
\label{Pdz_75}
\alpha = \sigma sin\Theta_{W} + \beta cos\Theta_{W} = 0 \; .
\end{equation}

Now we have the result that the ground fields in  Eq.(\ref{Pdz_21})  can
be rewritten as follows:
\begin{eqnarray}
\label{Pdz_76}
\left\{ \begin{array}{lll}
W^{1,2}_{\mu} = 0 \;\; , \;\; W^{3}_{i} = 0 \; , \\
W^{3}_{0} = - \beta ctg\Theta_{W}  \; , \\
B_{0} = \beta  \; , \\
B_{i} = 0 \; , \\
\varphi = \delta \; .
\end{array} \right.
\end{eqnarray}
or in terms of physical ground fields
\begin{eqnarray}
\label{Pdz_77}
\left\{ \begin{array}{lll}
W^{\pm}_{\mu} = 0 \;\; , \;\; Z_{i} = 0 \; ,
\\
Z_{0} = \zeta \;\;\;\;\;  {\rm where} \;\;\;
(\zeta = -\frac{1}{sin\Theta_{W}} \; \beta)  \; , \\
A_{\mu} = 0  \; , \\
\varphi = \delta \; .
\end{array} \right.
\end{eqnarray}

The appearance of the non-zero value of the weak charge density
fluctuation $\varrho_{Z^{0} SM}$ and $\zeta$ boson ground state
field induced by it (see Eq.(\ref{Pdz_72})) influences the masses
of the fields in the model and from
Eqs.(\ref{Pdz_44})-(\ref{Pdz_47}) ($\vartheta = 0$ and $\alpha =
0$) we obtain (see Figs.12-13):
\begin{equation}
\label{Pdz_78}
m_{W^{\pm}}^{2} = \frac{1}{4}g^{2}\delta^{2} - g^{2} \zeta^{2}
cos^{2} \Theta_{W} \; ,
\end{equation}
\begin{equation}
\label{Pdz_79}
m_{Z^{0}}^{2} = \frac{1}{4}(g^{2} + g'^{2})\delta^{2} \; ,
\end{equation}
\begin{equation}
\label{Pdz_80}
m_{A}^{2} = 0 \; ,
\end{equation}
\begin{equation}
\label{Pdz_81}
\Delta_{\varphi}^2 = {\cal N} \times f(\zeta, \delta) \; .
%\left[ 2 \lambda (v^{2} + 3\delta^{2})
%- \frac{1}{4}(g^{2} + g'^{2})\zeta^{2} \right] \; .
\end{equation}
Let us notice that the effective mass of the physical field $A_{\mu}$ is
$m_{A}^{2} = 0$.

The minimal energy density of the bgfms configuration
${\cal E}_{min}(\varrho_{Z^{0} \; SM}) = {\cal U}_{ef} (\vartheta = 0,
\; \delta \neq 0 ) $ is (see Figure 14):
\begin{eqnarray}
\label{Pdz_82}
{\cal E}_{min }(\varrho_{Z^{0} \; SM})  =  \frac{1}{2}\zeta \sqrt{g^{2} +
g'^{2}} \varrho_{Z^{0} \; SM} + \frac{1}{64 \lambda}(g^{2} +
g'^{2})^{2}\zeta^{4} \; .
\end{eqnarray}

From the Eq.(\ref{Pdz_78}) it is clear that the appearance of
$\varrho_{Z^{0} SM} > 0$ (so the boson ground field $\zeta > 0$)
leads to the instability in the $W^{\pm}_{\mu}$ sector if
\begin{equation}
\label{Pdz_83}
\zeta^{3} (\varrho_{Z^{0} SM}) > \frac{\sqrt{g^{2} + g'^{2}}}{g^{2}}
\varrho_{Z^{0} \; SM} \; .
\end{equation}
When the equality $\zeta^{3} (\varrho_{Z^{0} SM}) = \varrho_{Z^{0} \; SM}
\sqrt{g^2 + g'^2}/g^{2}$ is
taken into account we obtain the relationship between $\lambda_{max}$ and
$\varrho_{Z^{0} \; max}$, where  $\lambda_{max}$ is the value of $\lambda$ and
$\varrho_{Z^{0} \; max}$ is the value of $\varrho_{Z^{0} \; SM}$ for which we
have $m_{W^{\pm}}^{2} = 0$ (see Figure 15).
The region of possible bgfms configurations with $\zeta \neq 0$ is on and
below the $\lambda_{max}-\varrho_{Z^{0} \; max}$ curve.

For weak charge density fluctuation $\varrho_{Z^{0} \; SM} \leq g
v^{3}/ (8 cos^{2} \Theta_{W}) \approx 1.655 \; 10^{6} \; $
$GeV^{3}$ this configuration of fields is stable for an arbitrary
$\lambda$ (see Figure 15). For values of $\varrho_{Z^{0} \; SM}$
bigger than $1.655 \; 10^{6} \; GeV^{3}$, this configuration of
fields for given $\lambda$ will be destabilized at certain value
of $\varrho_{Z^{0} \; SM} = \varrho_{Z^{0} \; max}$ and the
system could reach the charged (with $\varrho_{Q \; SM} \neq 0$)
stable configuration of fields with $\vartheta \neq 0$. For
$\lambda < g^{2}/(16 \; cos^{4}\Theta_{W}) \approx 0.0422 $ the
configuration of fields is stable for all values of weak charge
density fluctuations $\varrho_{Z^{0} \; SM}$ (see Figure 15).

We can also examine the mass $M_{i^3} = 4/3 \;\; \pi \; r_{i^3}^{3} \;
{\cal E}_{min}(\varrho_{Z^{0} \; SM})$ (see Figure 16) of a bgfms
configuration with non-zero weak charge density fluctuation.
Here $r_{i^3}$ is the "radius of
the charge fluctuation" of this configuration which has the matter global weak
isotopic charge fluctuation $i^{3} = 4/3 \;\; \pi \; r_{i^3}^{3} \;
\varrho_{Z^{0} \; SM}$.

Taking into account the mass of a fermion (fermions), playing the
role of matter source inducing boson ground fields, will change
the value of the mass $M_{i^3}$ of about the order of the mass of
this part of a fermion (fermions) which is contained in the
region of the fluctuation. These configurations of fields lie
only on the $M_{i^3} -  r_{i^3}$ curve, where $M_{i^3} = \pm i^3
\times M_{i^3 = 1}$ (sign "+" for $I^{3} > 0$, sign "-" for
$I^{3} < 0$). The function $M_{i^3 = 1} (r_{i^3})$ is presented
in Figure 16. In the case of neutron (or neutrino) its mass
$m_{n}$\footnote{ It is energeticly more favorable for the
fluctuation to appear in the whole region of neutron ($r_{i^3} =
r_{Z}$) than in its part (our model is to simple to prove it). }
(or $m_{\nu}$) corrects the mass $M_{i^3}$ of the droplet by the
value of $\sim m_{n}$ (or $m_{\nu}$). Hence according to Figure 16
the mass of the droplet lies on the curve which gives its mass
which is slightly above the mass $m_{n}$. The mass of the droplet
($\sim m_{n} + \sim 1 \; keV$) lies almost in the region of
uncertainty of the neutron mass $m_{n} \approx 0.9396 \; GeV$. In
the case of a neutrino the mass of a connected droplet might be
even bigger depending on its radius (which is presumed to be very
small). But the most important fact is that the physical ground
fields $\delta$ and $\zeta$ (see Eq.(\ref{Pdz_77})) are present
in the droplet region - the droplet is the configuration of
fields. Because of this the droplet cannot decay to neutron and
photon unless the energy equal to the sum of the mass of the $Z$
particle and the value of $|\Delta_{\varphi}|$ outside the
droplet is supplied to the droplet.

We can also obtain the upper (according to the stability of this
configuration of fields within the $W^{\pm}$ sector) limit $M_{i^3, max}$
for the value of the mass $M_{i^3}$ with the region of possible bgfms
configurations which lie on and below the $M_{i^3, max}-\lambda_{max}$
curve (see Figure 17).

With the experimental knowledge of the mass $M_{i^3, max} + \,
\sim$ mass of a host fermion, and using this curve (see Figure 17)
the value of $\lambda$ can be calculated. The function
$M_{i^3=1}(r_{i^{3}})$ is also presented in Figure 18.

In conclusion, when extended $I^{3} \neq 0$ and the weak charge
density fluctuation is non-zero $\varrho_{Z^{0} \; SM} \neq 0$
(but $\varrho_{Q \; SM} = 0$) then the field configuration with
$Z_{\mu}$ gauge ground field exists. The asymptotic case
$\varrho_{Z^{0} \; SM} \longrightarrow 0$ produces the
electroweak assumptions (extended $I^{3} = 0$) with  and $\delta
= v$.

\subsection{The droplet and the process of pair production
$\gamma + \gamma \rightarrow e^{+} + e^{-}$}
\label{process}

Now very important fact should be noticed.
It is clear that in droplets of bosonic ground state fields induced by
$\varrho_{Z^{0} \; SM} \neq 0$ (but $\varrho_{Q \; SM} = 0$) the conversion
$\gamma + \gamma \rightarrow e^{+} + e^{-}$ is not allowed.
The reason is very simple.
Let us imagine that an electrically charged particle (such as $e^{+}$ or
$e^{-}$) appears inside the droplet of this configuration of fields.
Because it has non-zero value of $Q$ (in the extended form), it should give
rise to the appearance (see Section~{\bf \ref{electric}},
Figure 10) of a droplet with the electric charge $Q =$
($1 + $ charge fluctuation $q$) for
$e^{+}$ (or $Q =$ ($-1$ + charge fluctuation $q$) for $e^{-}$) and the mass
equal at least to $M_{q} \approx q \times 80.13 \; GeV$.
Now we recall that inside the initial droplet with $\varrho_{Z^{0} \; SM}
\neq 0$ (but $\varrho_{Q \; SM} = 0$) a photon has the effective mass equal to
zero (see Eq.(\ref{Pdz_80})).
Hence the droplet with $\varrho_{Z^{0} \; SM} \neq 0$ (but $\varrho_{Q \; SM}
= 0$) is transparent for photons observed in gamma-ray bursts
\cite{JankaRuffert}. \label{bursts1}

\section{Conclusions}
\label{boson-concl}

In the present paper the "non-linear" self-consistent theory of
classical fields in the electroweak model has been proposed.
Homogeneous boson ground state solutions in the GSW model at the
presence of a non-zero extended fermionic charge densities have
been reviewed and fully reinterpreted to make the theory with
non-zero charge densities \cite{JacekManka} sound. Consequences of
charge density fluctuations are proposed \cite{MarekJacek}.

But in order to understand the model in the broader context
let us for a moment simplify  our considerations taking into account
a real scalar field theory model defined by the
following Lagrangian density:
\begin{eqnarray}
\label{scalar-Lagr} {\cal L} = \frac{1}{2} \dot{\phi}^{2}(x,t) -
\frac{1}{2}(\nabla \phi(x,t))^{2} - V(\phi)
\end{eqnarray}
where $\dot{\phi} = \partial \phi/\partial t$, $\nabla \phi =
\sum_{i} \bar{i} \; \partial\phi/\partial x^{i}$ ($\bar{i}$ is
the versor).

$V(\phi)$ is a function of $\phi$ and the dependence on the
coupling constant $g$ is given by
\begin{eqnarray}
\label{V-phi} V(\phi) = \frac{1}{g^{2}} \tilde{V}(\chi), \; \; \;
\chi = g \phi
\end{eqnarray}
where $\tilde{V}$ is an even function which is independent of $g$.
Depending on the choice of the functional form of $\tilde{V}$ one
can consider various models. One of its realization is presented
on Figure 19.

The Hamiltonian density derived from the Lagrangian of
Eq.(\ref{scalar-Lagr}) is given by
\begin{eqnarray}
\label{scalar-Hamilt} {\cal H} = \frac{1}{2} \dot{\phi}^{2} +
\frac{1}{2}(\nabla \phi)^{2} + V(\phi)
\end{eqnarray}
Let $\phi_{0}$ be the scalar timeless field solution of the
equation of motion for the field $\phi$ in the ground state of
the system given by this Hamiltonian. I use the name of a {\it
scalar ground field} for the solution $\phi_{0}$.

When we are interested in the Lagrangian density
\begin{equation}
\label{lagr-el} {\cal L} = \bar{\Psi} ( \gamma^{\mu} i
\partial_{\mu} - m ) \Psi + J^{\mu} \, A_{\mu} - \frac{1}{4} \,
F_{\mu \nu} \, F^{\mu \nu} \, ,
\end{equation}
where $J^{\mu} = - e \bar{\Psi} \gamma^{\mu} \Psi $ is the
electron current density fluctuation and $A_{\mu} $ is the total
electromagnetic field, four-potential $A_{\mu} = A^{e}_{\mu} +
A^{s}_{\mu} $, with the superscripts $e$ and $s$ standing for
external field and self field (which is adjusted by the radiative
reaction to suit the electron current and its fluctuations, see
\cite{Barut}), respectively, then, in the minimum of the
corresponding total Hamiltonian, the solution of the equation of
motion for $A^{s}_{\mu} $ is called {\it electromagnetic ground
field}.

More generally we have used the name of a {\it boson ground field}
for a solution of an equation of motion for a {\it boson field}
in the ground state of a whole system of fields (fermion, gauge
boson, scalar) under consideration. This boson field is the self
field  (or can be treated like this) when it is coupled to a
source-"basic" field
%described by the wave mechanical wave function
. By "basic" field we have meant a wave (field) function which is
proper for a fermion, a scalar or a heavy boson. This concept of
a wave function and the Schr\"{o}dinger wave equation is dominant
in the nonrelativistic physics of atoms, molecules and condensed
matter. In the relativistic quantum theory this notion had been
largely abandoned in favour of the second quantized perturbative
Feynman graph approach, although the Dirac wave equation is used
approximatively in some problems. What was done by Barut and
others was an extension of the Schr\"{o}dinger's "charge density
interpretation" of a wave function\footnote{ Electron is a
classical distribution of charge.} to a "fully-fledged"
relativistic theory. They implemented successfully this "natural
(fields theory) interpretation" of a wave function in many
specific problems with coupled Dirac and Maxwell equations (for
characteristic boundary conditions). But the "natural
interpretation" of the wave function could be extended to the
Klein-Gordon equation \cite{Dziekuje Ci Panie Jezu Chryste}
coupled to Einstein field equations, thus being a rival for
quantum gravity in its second quantization form. In both cases the
second quantization approach is connected with the probabilistic
interpretation of quantum theory, whereas the "natural
interpretation" together with the self field concept goes in tune
with the deterministic interpretation composing a relativistic,
self-consistent field theory.

To summarize: Depending on the model, the role of a self field
can be played by electromagnetic field \cite{bib B-K-1,bib
B-H,bib B-D,bib B,spontaneous,Unal,bib J,bib M}, boson
$W^{+}-W^{-}$ ground-field (this paper, for example), or by the
gravitational field (metric tensor) $g_{\mu \nu}$ \cite{Dziekuje
Ci Panie Jezu Chryste}. The main law for arising of these self
fields would be taking the lead existence of "basic" fields.

Now, in view of this language, let us conclude what have been
done in the present paper. In the presence of the external matter
sources boson Higgs ground field and gauge ground fields were
examined. In general, we notice two physically different
configurations of fields. When the charge density fluctuation
$\varrho_{Q \; SM}$ is not equal to zero, then bgfms
(boson ground fields induced by matter sources) classes\footnote{
Crossing from one bgfms class (they are characterized by values
of ground fields) of configuration of fields to another class
which has different quantum number is somehow forbiden. So, also,
one bgfms droplet with certain values of quantum numbers does not
convert to another configuration of fields with different quantum
numbers. Hence, for example, an electrically neutral bgfms
droplet might not convert itself to one of its predecessors (to
neutron in neutron star - for example). } of configurations of
fields with $\vartheta \neq 0 $ and $\delta \neq 0$ exist. In
this configuration of fields the $W^{\pm}_{\mu}$ bosons are
massless while the electromagnetic fields $A_{\mu}$ and bosons
$Z_{\mu}$ are massive. The appearance of the mass of the
$A_{\mu}$ field is the result of the screening effect
\cite{Aitchison-Hey-bis}. We observe very deep energy density
minimum (see Figure 9) with ${\cal E}_{min} \approx 44.382 \;
GeV^{4}$ and the charge density fluctuation $\varrho_{Q \; SM}
\approx 0.5539 \; GeV^{3}$ for which we obtained a droplet of
this configuration of field with the "radius of the charge
fluctuation" $r_{q} \approx q^{1/3} \times 0.149 \; fm$ and the
mass (in the thin wall approximation) $M_{q} = \pm q \times 80.13
\; $ $GeV$ (for matter global electric charge fluctuation equal
to $q$). Hence droplets of this configuration of fields could be
experimentally observed by their very small ratio $|q|/M_{q} \leq
1/80.13 \; GeV^{-1}$. The mass of a droplet of this configuration
of fields $M_{q} \longrightarrow \pm q g v/2 = \pm q \times 80.13
\; GeV \;\; {\rm as} \;\; \varrho_{Q \; SM} \longrightarrow 0 $
for all values of $\lambda > 0$. These bgfms configurations lie
on the $M_{q}-r_{q}$ curve (see Figure 10) or equivalently on the
$E_{min}-\varrho_{Q \; SM}$ curve (see Figure 9) only.

When $\varrho_{Q \; SM} = 0$ and $\varrho_{Z^{0} \; SM}
\neq 0$ then the second configuration of fields ($\vartheta = 0$ and
$\zeta \neq 0$), with the $Z_{\mu}$ gauge ground field, exists.
When $\varrho_{Z^{0} \; SM} \longrightarrow 0$ then this case gives the
electroweak assumptions of the GSW model with $\delta = v$.
Now the region of possible bgfms configurations lies on and below the
$\lambda_{max}-\varrho_{Z^{0} \; max}$ curve (see Figure 15).
For $\lambda < g^{2}/(16 \; cos^{4}\Theta_{W}) \approx 0.0422 $, the
configuration of fields is stable for all values of the weak charge density
fluctuation $\varrho_{Z^{0} \; SM}$.
However, it is stable for an arbitrary $\lambda$ only if
$\varrho_{Z^{0} \; SM} \leq g v^{3}/(8 cos^{2}\Theta_{W}) \approx 1.655 \;
10^{6} \; GeV^{3}$.
For $\varrho_{Z^{0} \; SM} > \varrho_{Z^{0} \; max} \approx 1.655 \;
10^{6} GeV^{3}$ this configuration of fields, for given $\lambda$, will be
destabilized at certain value of
$\varrho_{Z^{0} \; SM} = \varrho_{Z^{0} \; max}$ and the
system could reach the charged (with $\varrho_{Q \; SM} \neq 0$) stable
configuration of fields with $\vartheta \neq 0$.

The further evolution of the system for $\varrho_{Q} \neq 0$ and $\vartheta
\neq 0$ seems to be very interesting.
It may lead, in the electromagnetic field outside of the
droplet of this bgfms configuration, to the particle-antiparticle pair
production. In effect, the bgfms droplet will gain lower charge by catching
one of them and the other will be sent. The obtained bgfms droplet will have
lower energy, reaching a metastable point for another bgfms configuration with
lower $\varrho_{Q}$ charge and so on until both $\varrho_{Q} = 0$ and
$\varrho_{Z^{0}} < \varrho_{Z^{0} \; max}$.

Field theories with point-like charges do not have such solutions and hence
do not give new possibilities to describe more complicated structures
(wavefunctions) of matter.

Are there new observations behind this fact? We can answer this
question in affirmative. The difference between the inward
structure of the neutron and the inward structure of a droplet
consisting of bosonic ground state fields appearing in the GSW
electroweak model\footnote{\label{bursts2}On this base an
alternative source of energy which may fertilize the gamma-ray
bursts was proposed \cite{MarekJacek}.} may be a supporting
impulse to start off the relativistic fireball in the collapsing
object (neutron star mergers) commonly believed to be responsible
for gamma-ray bursts.

Here, on earth, we perceived the appearance of such ground state boson field
configurations in very dense microscopic objects created in heavy ion
collisions \cite{bib Muller}. There is no necessity to repeat these
experiments but to reinterpret the results.

The discussed model is homogeneous on the level of one droplet.
Next calculations should incorporate more realistic shapes of the
charge densities of extended matter sources. These shapes would
follow from the coupled Klein-Gordon-Maxwell (Yang-Mills) or
Dirac-Maxwell (Yang-Mills) equations for charge density
fluctuations as it is required in the self-consistent models.
Sometimes even Einstein equations should be entangled\footnote{
Hence, I see a matter particle to be, from the mathematical
point of view, a self-consistent solution of field equations
involved in the description of this particle. }. Finally,
presented model is a step towards (the self field formalism of) a
classical theory of a one elementary particle which is a material,
extended entity, having electroweak, gravitational,  etc. self
fields of its own, coupled to it. So, next models shall be
concerned with a structure of one particle.

\newpage

\section{Table: Quantum numbers} \nopagebreak[4]
\label{boson-tabl}

\ \ \ \ Some quantum numbers in the $SU_{L}(2) \times U_{Y}(1)$ electroweak
theory.
\nopagebreak[4]

\begin{tabular}{ccccc}   \hline

           & Weak           & Weak         & Electric          &    \\
           & Isotopic       & Hypercharge  & Charge $Q$        &$p = 2Q/Y$ \\
           & Charge $I^{3}$ & $Y$          & $ Q = I^{3} +Y/2$ &     \\
\hline

$Leptons$         &         &              &                   &     \\

$\nu_{L}$         &    1/2  &  - 1         &    0              &   0 \\
$ e_{L}  $        &  - 1/2  &  - 1         &  - 1              &   2 \\
                  &         &              &                   &     \\
$e_{R}$           &     0   &  - 2         &  - 1              &   1 \\
\hline

$Gauge \; Bosons$ &         &              &                   &     \\

$W^{+}$           &     1   &    0         &    1              &     \\
$W^{3}$           &     0   &    0         &    0              &     \\
$W^{-}$           &  -  1   &    0         &  - 1              &     \\
                  &         &              &                   &     \\
$ B $             &     0   &    0         &    0              &     \\
\hline

$Higgs \; Boson$  &         &              &                   &     \\

$ H^{+} $         &    1/2  &    1         &    1              &   2 \\
$ H^{0}$          &  - 1/2  &    1         &    0              &   0 \\
\hline
                  &         &              &                   &          \\
$Some$            &  - 1/2  &    1         &    0              &   0      \\
$matter$          &   1/2   &    1         &    1              &   2      \\
$source$          &    0    &    2         &    1              &   1    \\
$configurations$  &  - 1    &    4         &    1              &   1/2  \\
                  &         &              &                   &        \\
\hline

\end{tabular}

\section*{Acknowledgments}
This work has been supported by S.B.S.

\newpage

\section*{Figure captions}

{\bf Figure 1}
%file: alfa_1.eps

The $\alpha$ ground field of the $A_{0}$ gauge boson fields as
the function of the standard electric charge density fluctuation
$\varrho_{Q \; SM }$($\vartheta \neq 0$, $\delta \neq 0$).

{\bf Figure 2}
%file: dzeta_1.eps

The $\zeta$ ground field of the $Z_{0}$ gauge boson fields as the
function of the standard electric charge density fluctuation
$\varrho_{Q \; SM }$($\vartheta \neq 0$, $\delta \neq 0$).

{\bf Figure 3}
%file: theta_1.eps

The $\vartheta$ ground field of the $W^{a=1}_{i=2}$ and
$W^{a=2}_{i=1}$ gauge boson fields as the function of the
standard electric charge density fluctuation $\varrho_{Q \; SM
}$($\vartheta \neq 0$, $\delta \neq 0$).

{\bf Figure 4}
%file: delta_1.eps

The $\delta$ ground field as the function of the standard
electric charge density $\varrho_{Q \; SM }$ ($\vartheta \neq 0$,
$\delta \neq 0$).

{\bf Figure 5}
%file: mz_1.eps

The mass $m_{Z}$ of the $Z^{\mu}$ gauge boson fields as the
function of the standard electric charge density fluctuation
$\varrho_{Q \; SM }$ ($\vartheta \neq 0$, $\delta \neq 0$).

{\bf Figure 6}
% file: ma_1.eps

The mass $m_{A}$ of the $A^{\mu}$ gauge boson fields as the
function of the standard electric charge density fluctuation
$\varrho_{Q \; SM }$ ($\vartheta \neq 0$, $\delta \neq 0$). The
region of $\varrho_{Q \; SM }$ which is in the range up to values
approximately $10^{3}$ times bigger than these for the matter
densities in nucleon matter is indicated by the arrow.

{\bf Figure 7}
% file: sinus_1.eps

The ratio $sin\Theta/sin\Theta_{W}$ (the $\Theta$ angle is the
modified mixing angle, see.Eq.(47)) as the function of the
standard electric charge density fluctuation $\varrho_{Q \; SM
}$($\vartheta \neq 0$, $\delta \neq 0$).

{\bf Figure 8}
% file: roq_1.eps

The physical electric charge density fluctuation $\varrho_{Q}$
(see Eq.(44)) as the function of the standard electric charge
density fluctuation $\varrho_{Q \; SM }$($\vartheta \neq 0$,
$\delta \neq 0$). The region of $\varrho_{Q \; SM }$ which is in
the range up to values approximately $10^{3}$ times bigger than
those for the matter densities in nucleon matter is indicated by
the arrow.

{\bf Figure 9}
%file: emin-l_1.eps

The~minimal~energy~density of~the~bgfms~configuration ${\cal
E}_{min}(\varrho_{Q \; SM})$ $= {\cal U}_{ef}(\vartheta \neq 0 ,
\; \delta \neq 0 \; )$ (see Eq.(24)) for boson ground fields given
by Eqs.(48)-(51) (for all values of $p \neq 0$ from Section~6) as
the function of the standard electric charge density fluctuation
$\varrho_{Q \; SM }$ which is in the range up to values
approximately $10^{3}$ times bigger than charge densities for
nucleon matter. The region in the vicinity of the "stable" bgfms
configuration (see Eq.(54)) is indicated by the arrow.

{\bf Figure 10}
% file: M-rq_1.eps

The mass $M_{q=1}$ of the bgfms configuration for boson ground
fields given by Eqs.(48)-(51) as the function of the "radius of
the charge fluctuation" $r_{q}$ ($\vartheta \neq 0$, $\delta \neq
0$ and for all values of $p \neq 0$ from Section~6). The region in
the vicinity of the "stable" bgfms configuration is indicated by
the arrow.

{\bf Figure 11}
% file: uef-d_2.eps

The self-consistent classical effective potential ${\cal
U}_{ef}(\delta; \;\; \vartheta = 0,\; \varrho_{Q \; SM} = 0)$ as
the function of the $\delta$ Higgs ground field ($\lambda=1$).

{\bf Figure 12}
% file: mw-roz_2.eps

The square mass $m_{W^{\pm}}^{2} $ of the $W^{\pm}_{\mu}$ gauge
boson fields (see Eq.(66)) as the function of the standard weak
charge density fluctuation $\varrho_{Z^{0} \; SM
}(\vartheta~=0,\delta~\neq~0, \varrho_{Q \; SM}=0$).

{\bf Figure 13}
% file: mz-roz_2.eps

The mass $m_{Z}$ of the $Z$ gauge boson fields (Eq.(67)) as the
function of the standard weak charge density fluctuation
$\varrho_{Z^{0} \, SM}(\delta \neq 0,\varrho_{Q \;
SM}=\vartheta=0$).

{\bf Figure 14}
% file: emin-l_2.eps

The minimal energy density of the bgfms configuration ${\cal
E}_{min }(\varrho_{Z^{0} \; SM}) = {\cal U}_{ef}(\vartheta=0,\;
\delta \neq 0, \; \varrho_{Q \; SM} = 0)$ (see Eq.(70)).

{\bf Figure 15}
% file: l-roz_2.eps

The partition of the ($\lambda,\varrho_{Z^{0} \; SM }$) plane
into two regions of stability and instability of the bgfms
configuration of fields with $\vartheta = 0$ and $\delta \neq 0$.
The region of possible bgfms configurations lies on and below the
$\lambda_{max}-\varrho_{Z^{0} \; max}$ curve where
$\lambda_{max}$ is the value of $\lambda$ and $\varrho_{Z^{0} \;
max}$ is the value of $\varrho_{Z^{0} \; SM}$ for which we have
$m_{W^{\pm}}^{2} = 0$.

{\bf Figure 16}
% file: Mi3-r_2.eps

The mass $M_{i^3 = - 1/2}$ of the bgfms configuration as the
function of the  "radius of the charge fluctuation" $r_{i^3}$.
Taking for granted that the radius $r_{i^3}$ of the fluctuation
is of $1 \, fm$ order we see that its mass is of $\sim m_{n} + 1
\, keV$ order (if the extended charge density is carried by
neutron with mass $m_{n}$).

{\bf Figure 17}
% file: Mi3-l_2.eps

The~upper~mass $M_{i^{3}=1, max}$ of a bgfms configuration of
fields with $\vartheta=0$ and $\delta \neq 0$ as the function of
the $\lambda_{max}$ non-linear Higgs parameter where
$\lambda_{max}$ and $M_{i^3=1, max}$  are  the  values   of
$\lambda$   and   $M_{i^3=1}$ respectively for which
$m_{W^{\pm}}^{2} = 0$ (on the curve). Here the region of possible
bgfms configurations of fields is on and below the
$\lambda_{max}-M_{i^3, max}$ curve.

{\bf Figure 18}
% file: Miup-r_2.eps

The upper (according to the stability of the bgfms configuration
with $\vartheta = 0$ and $\delta \neq 0$ in the ${W}^{\pm}$
sector) mass $M_{i^3=1, max}(r_{i^{3}})$ of a bgfms configuration
(with the weak isotopic charge $i^{3}=1$) as the function of the
"radius of the charge fluctuation" $r_{i^{3}}$ of this
configuration.

{\bf Figure 19}
% file: V-chi_c.eps

The potential $\tilde{V}(\chi$) for a toy model in scalar field
theory.

\end{document}